# CAN GRAPHENE BILAYERS BE THE MEMBRANE MIMETIC MATERIALS?

## Gradov Oleg V.

Institute of Energy Problems of Chemical Physics, Russian Academy of Sciences, http://www.inepcp.ru
38/2, Leninskii prosp., 117829 Moscow, Russian Federation
gradov@chph.ras.ru

Since the pioneering works of the founder of membrane mimetic chemistry Janos H. Fendler it is known that a number of atomic or molecular clusters and films (including nanoscale ones) are capable of mimicking the membrane functions. Membrane mimetic materials can be either soft matter or solid state materials. Conducting films (including those with magnetic properties) and semiconductors are also known to possess membrane mimetic properties. If we consider the agent exchange through the membrane in the operator form, the chemical composition of the membranes and their models, as well as the difference between the atomic and molecular clusters or layers become not so essential, and hence, membrane mimetic chemistry of nano- and mesostructures do not differ significantly within the agent-based approach. This invited review containing several parts reflects the main aspects of the author's report at the conference "Graphene": a molecule and 2D-crystal" (September 8-12, 2015, Novosibirsk, Russia) and considers various aspects of the similarity between the graphene nanostructures, membranes and bionic membrane-like nanomaterials.

*Keywords:* bilayer graphene, bilipid layers, membrane mimetics, nanobionics

PACS: 81.05.ue

## Contents

## 1. INTRODUCTION. A PROBLEM OF COMPLETENESS OF THE MEMBRANE MIMETIC DESIGN AND MODELING

Membrane structures are known to form the basis of operation of all the physiological (including cytophysiological) systems, providing selective transport of ions and other agents into the cytoplasm, maintenance of the biochemical gradients required for many physiological processes, molecular recognition/reception at the cellular surface, electrophysiological excitation/inhibition due to the biopotential generation and transduction, labile template fixation of the interacting membrane proteins, mechanical stability and the barrier function, intercellular interaction, reaction-diffusion morphogenetic processes changing the form of the organelles, cells and tissues up to the organs and an organism as a whole [1, 2].

Thus, a system or structure considered as an adequate membrane-mimicking material should ideally possess not less than a full number of the physico-chemical properties (or at least their phenomenological analogs) providing the performance of the above mentioned functions of the natural biological membranes. Otherwise, there is no reason to compare the functions of the model and natural membranes, since the physico-chemical systems capable of performing several function of the biomembranes (e.g. only ion exchange, energy storage or mechanical functions) are widespread both in nature and industry, but their consideration either as the reference or as the absolute membrane models leads to the reduction of the fundamentally important variables and introduction of the a priori normalizations/assumptions required for simplification and the inevitable error minimization, and decrease the heuristic value of the modeling itself.

As follows from the modeling theory, the criteria of the lover level model incompleteness are its impossibility to perform the functions of the higher level structures and the necessity to introduce additional components in order to compensate the lack of a certain function at the lower level [3-5]. This follows from the fact that for an adequate both structural and functional model should exist a one-to-one mapping mapping between the functional units of the original and the elementary units of the model. In the case of the artificial membrane models (membrane mimetic materials) the criterion of their insufficiency as the basic units of the higher level models is evident – it is the impossibility of most of them to perform a full number of the above mentioned functions characteristic for the biological cell membranes.

## 2. THE AMBIGUITY OF THE MEMBRANE MODEL DEFINITIONS: SEMI-SYNTHETIC, SYNTHETIC AND BIOGENIC MODEL MEMBRANES

This can be illustrated by multiple examples. Despite the active introduction of the membrane technologies into the "artificial cell" fabrication since the early 1960-th [6, 7], a complex number of the biomembrane functions has never been reproduced, because of using simplified membrane mimetic materials (including synthetic ones, such as nylon), with the properties rather far from those of the biological membranes and their structural intercalates. As a consequence, a widely accepted idea of the independence of the cell model properties from the properties of the membrane mimetic material used lead to the absence of an adequate complex cell model based on the membrane mimetic material without a full amount of the biomembrane functions. This is not surprising, since a hypothetic artificial cell wit such a membrane could not be able to perform all the necessary functions, including the metabolism, and the most adequate cell models are based on the bioorganic components, which can be considered as a kind of reduction rather than a cell modeling. Classical membrane mimetic materials in this case include a phospholipid bilayer [8,9], which is also a component of the biomembranes [10]; myelin – a component of the nerve fiber membrane containing a special protein – myelin basic protein (MBP), also represents a membrane mimetic material applied either as an emulsion or in a micellar form [11, 12]. However, it is evident that all the above substances are in fact the components of natural membranes, and hence, can not be considered as the true membrane mimetic materials. Even in the case of the membrane function modeling using the physico-chemical properties of the materials (e.g. polyelectrolyes), the main components include standard lipids available in the natural cells [13]. The main differences between the artificial and natural membranes preventing the membrane mimetic systems with an insufficient descriptor set from reproducing all the functions of the membrane compartments are listed in the review paper [14], so they can be omitted here.

In other words, there is a problem of re-interpretation of a "minimal cell" not as a cell with a minimal genome [15-21] and a corresponding nucleotide metabolism [22-24], but as a supramolecular structure built up with membranes with the properties which as a whole are necessary and sufficient for the operation of the functionally-minimal cell or its model. To date there are several papers considering the above problem, but the solutions proposed there imply conventional phospholipids already adapted to perform the functions simulated instead of the artificial membrane mimetic materials. Some authors [25, 26] use the well-known lipids in design of the minimal / artificial cell. Many papers emphasize the presence of the biogenic substances in the minimal cell, or even a «semi-synthetic» character of the artificial cell models proposed [27, 28] (synthetic aspect is just within the scope of biomimetics [29, 30]). Obviously, the model as an object of investigation in this case does not exactly correspond to the model as a subject of the studies, while the subject partly overlaps with another object under investigation. Thus, if a cell model or an artificial cell contains a fragment of the natural cell membrane, it can not be considered as an adequate synthetic cell model dye to its non-synthetic nature and biological origin.

## 3. THE DIFFERENCE BETWEEN MEMBRANE MIMETICS AND MEMBRANE BIOMIMETICS PHYSICAL PROPERTIES OF THE MEMBRANE MIMETIC MATERIALS IN THE SUBJECTIVE MODELS

In this connection there is a problem to distinguish between the membrane-mimetics and membrane biomimetics: a system can be a monoparametric physical membrane mimetic system, but not a membrane biomimetic system as a whole. The converse is not true. Thus, it is necessary to consider the basic principles of membrane-mimetics and determine their application limits to the biomembrane modeling. From the pioneering papers by Fendler [31, 32] who laid the foundation for "membrane mimetic chemistry" it is known that some atomic or molecular clusters and films (including nanoscale ones) are capable to mimic membrane functions. Membrane mimetic properties are not a unique feature of the "soft matter", but also can be found among the solid state carriers, including the immobilized ones [33]. According to the theoretical predictions and experimental data of electrochemistry and solid state physics, membrane mimetic properties can be manifested in both conductor and semiconductor films, including those with the magnetic properties [34-37]. The contradictory properties of different membrane mimetic materials indicate the arbitrary nature of the "similarity criteria" chosen, while the difference between the structural levels of similarity up to the membrane mimetic objects which differ even in their dimension from the biomembranes, such as gradient multilayer nanofilms made up with the quantum dots, low-dimensional superconductors, atomic monolayers or a two-dimensional electron gas, makes it impossible to define the membrane mimetic materials according to their basic physical processes and requires a transition to a higher level of abstraction, i.e. a formal mathematical consideration. Otherwise, a membrane could not be considered as an object of modeling and will become an object and subject of structural-functional speculations.

## 4. SEMICONDUCTOR MEMBRANE MIMETIC MATERIALS IN THE FRAMEWORK OF THE SOLID STATE PHYSICS AND THE EARLY MISCONCEPTIONS OF THE MEMBRANE THEORY

It is noteworthy that the contradictory approaches in membrane mimetic chemistry are similar to those in membrane biophysics and membrane biomimetics, which appeared at an early stage of their development in 1960-th - 1980-th. Thus, the uncertainty in the membrane-mimetic criteria is expected to disappear in the next years. In 1960-1980-th there were popular concepts about the semiconductor and "pseudo-solid-state" nature of the biomembrane functioning. It was believed that the neuron membranes possess semiconductivity [38] and the properties of an ionic psn-transition [39], as well as the properties of a semiconductor rectifier based on biomolecular lipid layers [40] (similar to the semiconductor pn-membranes [41]). There were also some papers considering a zone melting (or band recrystallization) as a model of an active transmembrane transport [42]. Semiconductor properties were also found in photoactive lipid membranes [43] and respiratory membrane proteins (terminal oxidase of an aerobic respiratory electron transport chain which catalyses the electron transfer to the oxygen molecular acceptor) [44]. At the same time there was an active development of the ideas inspired by the solid state physics (despite the fact that biomembranes belong to the soft matter [45]), which argued for the solid state character of a number of the fundamental biological processes. In particular, a pseudo-solid-state mechanism of the electron and ion transmembrane transport was proposed [46] (a so-called supramolecular solid state biology), but it was mostly speculative at that period due to the lack of the clear concepts describing the processes in polymeric solid state materials and supramolecular condensed associates, which appeared much later [47], especially for the charged membrane surfaces [48]. By that time there was a lack of data on the chiral solid state structures [49], solid carbohydrates [50] and ion-selective / ion-exchange structures [51], etc. which could be in some way associated with their biochemical prototypes. Some authors also argued for the hole (vacancy) mechanism of the ion competitve diffusion in the electrogenic neurocyte membranes [52], electron transfer in biostructures and bioenergetics based on this type of the charge transfer mechanism [53, 54], as well as the reactivity based on this bioenergetics. In particular, there was a solid state model of the well known Hodgkin-Huxley action potential generation and propagation mechanism [55], in general consistent with its conventional soft matter prototype). In early 1960-the there were also papers comparing a solid state energy transfer with the membrane activity during the photosynthesis processes [56]. This approach was intensively developed in the "era of the solid state and semiconductor physics" (until the 1990-th) before the period when the progress in molecular biology and the emergence of nanoscience allowed to understand the mechanism of electron transfer in biological systems in details. At the end of the 1980-th the words "solid state" in description of the operation mechanism of the electron transfer system elements (such as peroxidases – promoters in the soluble metalloprotein electrochemistry [57]) were used with the quotation marks (unlike the promoters for bacteria cultivation on solid media which do not require quotation marks [58, 59]). Therefore it can be assumed that the choice of the biomimetic material for the biomembrane modeling and the choice of the corresponding membrane models at that period were in accordance with the "scientific fashion" similar to the rapid shift of the synthetic membrane design towards the nanomaterial science as a result of the development of nanotechnology [60].

## 5. SUPERCONDUCTING MEMBRANE MIMETIC MATERIALS AND MEMBRANE MODELS

A similar example considers another membrane mimetic material from the Fendler list which also appeared in 1970-1980-th as a result of the scientific fashion – a film superconductor [36]. Although biomembrane structures do not possess superconductivity, there was a number of papers which attributed a sensitivity of some biological structures to the external fields and stimuli (magnetic field [61]; microwave radiation [62]; any affecters with the response described well with the Weber-Fechner law [63]) to superconductivity. In response to the reasonable questions about the reliance of such interpretation under physiological temperatures far from the cryogenic ones (which was in fact the introduction of the redundant entities into the description of the simple phenomena, well described within the existing concepts), the authors usually referred to the superconductivity observed in some organic / bioorganic systems at room temperature [64-66], which used to be rather popular in the early 1980-th when the papers in this area were published in the leading scientific journals [67-69], but nevertheless, the data presented required a careful separation of the facts from the artifacts, fantasy and misinterpretations [70]. This modern trend is in line with the similar periodically emerging concepts which lack strong physical basis, such as pseudo-scientific claims on the role of superconductivity in development of cancer [71] or in human health as a whole [72]. It is evident that membrane mimetic chemistry based on such incorrect principles will never be able to mimic a complex of biomembrane functions. Thus, both biophysical and membrane mimetic value of such approaches considering the similarity between the biomembranes and superconductive materials is rather doubtful, and it is inexpedient to discuss them here.

## 6. FERROELECTRIC MEMBRANE MODELS AND MEMBRANE MIMETIC MATERIALS

Another type of membrane mimetic materials – ferroelectric films, is more suitable for complex membrane mimetic modeling due to the presence of ferroelectric properties in bioorganic structures [73-76] and especially of the membrane surfaces. In the early 1990-th a simple and electrophysiologically-relevant model of the membrane excitation was proposed based on the analogy between the lipid membranes and diffuse ferroelectric bilayers performing the function of a "capacitor" similar to the bilayer membranes according to the early electrobiological concepts [77]. At the same time it was proposed to attribute the dielectric anomalies of the model biomembranes (such as a squid axon) to the temperature effects obeying the Curie-Weiss law which usually describes magnetic susceptibility of ferromagnetic materials in the paramagnetic zone above the Curie point (except the cases when the mean-field approximation is inapplicable) [78]. It is noteworthy that cytological and electrophysiological data confirm the membrane excitation interpretation as a response of the diffuse bilayers (like the Gouy layer) in many cases, including the ferroelectric bilayers [79]. For the models with a simple physico-chemical modeling of the prototype functions the chemical composition of the substrate does not matter, in contrast to the phase, rheological and other parameters. Therefore ferroelectric phenomena, similar to those applied in the lipid membrane models, also work in natural peptide nanotubes [80, 81] with an active surface. At a higher level of abstraction which follows from the above considerations, similar ferroelectric phenomena are typical for the polymer-based Langmuir-Blogett films [82], while the latter are known to mimic synaptosomal membrane structures [83] and biomembranes [84] with electrochemical activity [85] and membrane affinity [86], which is interpreted either as the membrane adhesion measured using a tensiometer, or as a conjugative ability studied by means of coordination or supramolecular chemistry, or even as the immunochemical affinity characterized by immunofluorescent and immunoelectrophoretic methods. Thus, ferroelectric structures possess a number of functional characteristics providing their applicability in the active membrane modeling (an incorrect term "segnetoelectric" frequently used in Russian language periodicals is synonymous to the term "ferroelectric" [87, 88]).

Functional similarity should be based on the strong structural or physico-chemical reason to be considered. Ion channels in biological cell membranes are responsible for the energy and mass transfer (active transport) and the electric signal transduction. Therefore, an adequate cell or biomembrane model should contain either membrane ion channels, which is impossible due to the difference in the operation conditions and artificial nature of the model, or their functional equivalent.

In the case of the synthetic ferroelectric membranes only a functional analogy can be realized without the substrate modeling, since the chemistry of the ion channel operation can not operate in the rheological conditions typical for the ferroelectric films.

Even a more sophisticated approach to the membrane function modeling does not require introduction of the special ion channels into the model membrane: in an ideal case the modeling medium itself should possess the properties sufficient for the ion channel function simulation. In 1987 there were two papers published in the same issue of the «Journal of Theoretical Biology», one of which postulated the existence of the special ferroelectric zones in a biomembrane channelome [89], while the second one states that the ion currents and the corresponding phase transitions in the cortical zone or cell membrane can be successfully described by the model ferroelectric channel units [90]. In 1992 Bystrov proposed a novel ferroelectric phason model of the membrane sodium ion channels [91] based on his early works on ferroelectrics with semiconductor properties where the ferroelectric response is realized via fluctuon and phason mechanisms [92]. This approach is in consistence with the above cited semiconductor membrane models. The signal transduction mechanism based on quasiparticles is characterized by a higher level of physical abstraction then the models using certain ion types which adequately mimic the biological prototype only at the specific biochemical / bioorganic medium. Later Bystrov moved from the sodium channel simulation [93] to the analysis of the ion channels [94] as the responsive structures with the gating phenomenon [95]. In the early papers [89,90] Leuchtag connected model concepts about bioferroelectric phenomena and the functions of the electrically-controlled voltage-dependent ion channels [96]. The striction effects observed were found to correlate well with the conformational and steric changes and the induced phase transitions in the ion channel structure and the surrounding membrane in the voltage-dependent ion channels [97, 98]. The only disadvantage of the model was the need to use the analogy with the superionic conductivity and the ferroelectric transition to the superionic state (the so-called superionic transition) in some relevant cases [97, 99], which caused objections from the specialists who worked with membranes beyond this transitional range. However, superionic phase transition is known in many cases to be a ferroelastic one and is characterized by a high probability of spontaneous deformation [100, 101], which allows its participation in ultrastructural morphogenesis in the presence of this effect in subcellular membranous structures. Speaking about the current trend considering mechanotransduction as a factor of development and morphogenesis [102], it is necessary to mention the concepts considering ferrofluids in morphogenesis modeling (for example, see «Morphogenesis: Origins of Patterns and Shapes», part II, «Ferrofluids: a model system of self-organized equilibrium» [103]). It is also noteworthy that a ferrofluid can be not ferromagnetic, but ferroelectric [104-107], while conventional ferrofluids or so-called ferromagnetic fluids, which do not retain residual magnetization in the absence of the external field in fact are not ferromagnetic, but rather paramagnetic or even superparamagnetic [108, 109]. They can even be neither liquids, nor solutions, but suspensions. An example of a non-liquid non-ferrofluid is the dusty space plasma with the properties of a superparamagnetic liqiud [110]). Ferrofluid application in tracing / self-assembly of the charge or charge-coupled flows (magnetoelecrostatic jets [109], etc) can be used in the models of self-organization under the membrane excitation of the ion channels based on the methods and principles of the ferroelectric physics [111]. Thus, it is possible to implement a higher integration level of the ion channel mimetics into the morphophysiological specialization of the biomodels using ferroelectric membranes and membranemimetic ferromagnetic structures, despite the fact that the mimicking medium does not possess the biochemical similarity and affinity to the biomembranes. In fact, this similarity ends at the level of the models which postulate the similarity between the ion transmembrane transport and the interfacial electron transport, which date back to the 1960-th [112]. However, the mimicking system in this case is more functional due to the presence of the physical analogs of the ion channels.

Therefore, similar functional similarity criteria can be applied for the analysis of the possibility to consider the graphene layers (particularly graphene bilayers) as the membrane mimetic materials and biomembrane models. This analysis will be presented in the next part of the paper considering the issues listed in the ansatz below.

## 7. ANSÄTZE

To perform the biomembrane mimetic functions graphene, as well as any other material, should possess:

1. Semipermeability, providing selective ion transport through the cell surface, and hence, the ionic channels/pores which are blocked by the heavy metals.

2. An electric double layer for energy accumulation (as in the capacitor), directed charge carrier gradient (such as the difference in electrochemical potentials of the protons at the coupling membranes in chemiosmotic theory introduced by P.D. Mitchell, leading to the charge accumulation at the membrane surface) and a periodic "capacity discharge".

3. The diffusion mechanism for the redox agents, such as water and gases in the simplest case of the passive transport, as well as the redox regulation of the osmotic liquid permeability; the cyclic exchange processes with the environment should cause the changes in the optical and electrophysical parameters of the chemical system surrounded by the membrane mimetic material.

4. Electrophysical activity, simulating or substituting the membrane potential generation as a dynamic process with a certain spectrum which can be considered as a signal transduction from the pseudo-membrane and induces a number of physico-chemical processes; the above process (especially in the case of the above cited ferroelectric membrane models with the conformation regulation by the external field) should be accompanied by the surface oscillations or a synchronous variation of the parameters correlated with the locally registered charge, such as a harmonic channel response.

5. The ability to form closed surfaces providing compartmentalization and gradient separation of the reaction processes, including those responsible for the electrogenesis, as well as the ability to duplicate the surface after reaching the critical value of the surface tension (or the critical mass) of the compartmentalized medium, similar to the simplest models of cytotomy.

6. Sensing properties to the primary physical stimuli with an unconditioned response shifting the equilibrium due to the primary charge transfer, which does not depend on the chemical composition of the surrounding medium.

7. Sensitivity towards the external chemical agents, shifting the equilibrium of the subsurface processes mediated by the membrane mimetic surface.

8. Sorption capacity (at least in the framework of the well-known simplified Roginsky-Zeldovich or Elovich equation) and on-surface template fixation of the separated chemical agents.

9. The ability to fix (either covalently or not) xenogenic sensor agents similar to the immune complexes grafted to the model phospholipid vesicles; such agents may include enzymes or other compounds capable of specific complementary supramolecular binding; in an ideal case the substitution of the enzyme and receptor functions should be performed without the enzyme or receptor molecules.

In the next part of this paper considering the above requirements to the graphene, graphene-like structures and, in particular, multilayered graphene, the section numbers correspond to those in the above list.

## REFERENCES

1. Luckey M. *Membrane Structural Biology: With Biochemical and Biophysical Foundations.* Cambridge University Press, 2008, 344 p.

2. Yeagle PL. *The Membranes of Cells.* Amsterdam, Academic Press (Elsevier), 2016, 452 p.

3. Kron G. *Diakoptics; the piecewise solution of large-scale systems.* London, MacDonald, 1963, 166 p.

4. Peschel M. *Modellbildung fur signale und systeme.* Berlin, VEB Verlag Technik,1978, 183 p.

5. Namatame A, Kurihara S, Nakashima H. (Eds.)*Emergent Intelligence of Networked Agents.* Berlin – Heidelberg, Springer, 2010, 258 p.

6. Chang TMS. *Artificial cells.* Springfield, Thomas, 1972, 207 p.

7.  Chang TMS. *Artificial cells: biotechnology, nanomedicine, regenerative medicine, blood substitutes, bioencapsulation, and cell/stem cell therapy.* New Jersey, World Scientific, 2007, 455 p.

8.  Bangham AD. Membrane models with phospholipids. *Prog. Biophys. Mol. Biol.*, 1968, 18: 29-95.

9.  Alkaitis D, Merola AJ, Lehninger AL. Phospholipid bilayers as biological membrane models: the effect of N,N'-bis(dichloroacetyl)-1,12-diaminododecane. *J. Membr. Biol.,* 1972, 10(3):237-246.

10. Thompson TE. Experimental bilayer membrane models. *Protoplasma*, 1967, 63(1):194-196.

11. Nicot C, Vacher M, Vincent M, Gallay J, Waks M. Membrane proteins in reverse micelles: myelin basic protein in a membrane-mimetic environment. *Biochemistry,* 1985, 24(24):7024-7032.

12. Chatenay D, Urbach W, Cazabat AM, Vacher M, Waks M. Proteins in membrane mimetic systems. Insertion of myelin basic protein into microemulsion droplets. *Biophys J.*, 1985, 48(6):893-898.

13. Tiourina OP, Radtchenko I, Sukhorukov GB, Möhwald H. Artificial cell based on lipid hollow polyelectrolyte microcapsules: channel reconstruction and membrane potential measurement. *J. Membr. Biol.*, 2002, 190(1):9-16.

14. Timashev SF. From Biological to Synthetic Membranes. *Russ. Chem. Bull.*, 1988, 57(6):487-503

15. Maniloff J. The minimal cell genome: "on being the right size". *Proc. Nat. Acad. Sci. USA.*, 1996, 93(19):10004-10006.

16. Islas S, Becerra A, Luisi PL, Lazcano A. Comparative genomics and the gene complement of a minimal cell. *Orig. Life Evol. Biosph.*, 2004, 34(1-2):243-256.

17. Bork P, Ouzounis C, Casari G, Schneider R, Sander C, Dolan M, Gilbert W, Gillevet PM. Exploring the Mycoplasma capricolum genome: a minimal cell reveals its physiology. *Mol. Microbiol.*, 1995, 16(5):955-967.

18. Browning ST, Castellanos M, Shuler ML. Robust control of initiation of prokaryotic chromosome replication: essential considerations for aminimal cell. *Biotech. Bioeng.*, 2004, 88(5):575-584.

19. Delaye L, Moya A. Evolution of reduced prokaryotic genomes and the minimal cell concept: variations on a theme. *Bioessays*, 2010, 32(4):281-287.

20. Zhang LY, Chang SH, Wang J. How to make a minimal genome for synthetic minimal cell. *Prot. Cell.*, 2010, 1(5):427-434.

21. Juhas M. On the road to synthetic life: the minimal cell and genome-scale engineering. *Crit Rev. Biotechnol.*, 2016, 36(3):416-423.

22. Castellanos M, Wilson DB, Shuler ML. A modular minimal cell model: purine and pyrimidine transport and metabolism. *Proc. Nat. Acad. Sci. USA.*, 2004, 101(17):6681-6686.

23. Oberholzer T, Wick R, Luisi PL, Biebricher CK. Enzymatic RNA replication in self-reproducing vesicles: an approach to a minimal cell. *Biochem. Biophys. Res. Commun.*, 1995, 207(1):250-257.

24. Lluch-Senar M, Delgado J, Chen WH, Lloréns-Rico V, O'Reilly FJ, Wodke JA, Unal EB, Yus E, Martínez S, Nichols RJ, Ferrar T, Vivancos A, Schmeisky A, Stülke J, van Noort V, Gavin AC, Bork P, Serrano L. Defining a minimal cell: essentiality of small ORFs and ncRNAs in a genome-reduced bacterium. *Mol. Syst. Biol.*, 2015, 11(1):780.

25. Murtas G. Internal lipid synthesis and vesicle growth as a step toward self-reproduction of the minimal cell. *Syst. Synth. Biol.*, 2010, 4(2):85-93.

26. Castellanos M, Kushiro K, Lai SK, Shuler

ML. A genomically/chemically complete module for synthesis of lipid membrane in a minimal cell. *Biotech. Bioeng.*, 2007, 97(2):397-409.

27. Murtas G. Question 7: construction of a semi-synthetic minimal cell: a model for early living cells. Orig. *Life Evol. Biosph.*, 2007, 37(4-5):419-22.

28. Caschera F, Noireaux V. Integration of biological parts toward the synthesis of a minimal cell. *Curr. Opin. Chem. Biol.*, 2014, 22:85-91.

29. Munteanu A, Solé RV. Phenotypic diversity and chaos in a minimal cell model. *J. Theor. Biol.*, 2006, 240(3):434-442.

30. Luisi PL, Stano P. Synthetic biology: minimal cell mimicry. *Nat. Chem.*, 2011, 3(10):755-756.

31. Fendler JH. Atomic and molecular clusters in membrane mimetic chemistry. *Chem. Rev,* 1987, 87(5):877–899.

32. Fendler J.H. Membrane-Mimetic Approach to Nanotechnology. *"Advances in the Applications of Membrane-Mimetic Chemistry"*. New York, Springer, 1994, pp. 1-15.

33. Pidgeon C. Solid phase membrane mimetics: immobilized artificial membranes. *Enz. Micr. Tech.*, 1990, 12(2):149-150.

34. Fendler JH. Metallic and catalytic particles. *Adv. Polym. Sci.*, 1994, 113:96-118.

35. Fendler JH. Semiconductor particles and particulate films. *Adv. Polym. Sci.*, 1994, 113:118-159.

36. Fendler JH. Conductors and superconductors. *Adv. Polym. Sci.*, 1994, 113:159-171.

37. Fendler JH. Magnetism, magnetic particles, and magnetic particulate films in membrane-mimetic compartments. *Adv. Polym. Sci.*, 1994, 113:172-181.

38. Cope FW. Evidence for semiconduction in Aplysia nerve membrane. *Proc. Nat. Acad. Sci. USA.* 1968. 61(3):905-908.

39. Adam G. Electrical characteristics of the ionic psn-junction as a model of the resting axon membrane. *J. Membr. Biol.*, 1970, 3(1):291-312.

40. Rosenberg B, Pant HC. The semiconducting rectifier behaviour of a bimolecular lipid membrane. *Chem. Phys. Lipids.*, 1970, 4(2):203-207.

41. Gracheva ME, Vidal J, Leburton JP. p-n Semiconductor membrane for electrically tunable ion current rectification and filtering. *Nano Lett.*, 2007, 7(6):1717-1722.

42. Dragomir CT. Zone melting as a model for active transport across the cell membrane. *J. Theor. Biol.*, 1971, 31(3):453-468.

43. Tien HT. Semiconducting Photoactive Bilayer Lipid Membranes. *Solut. Behav. Surfact.*, 1982, 1:229-240.

44. Cope FW. Semiconduction as the mechanism of the cytochrome oxidase reaction. Low activation energy of semiconduction measured for cytochrome oxidase protein. Solid state theory of cytochrome oxidase predicts observed kinetic peculiarities. *Physiol. Chem. Phys.*, 1979, 11(3):261-262.

45. Kleman M., Lavrentovich OD. *Soft matter physics*. New York, Springer, 2003, 637 p.

46. Cope FW. Supramolecular biology: a solid state physical approach to ion and electron transport. *Ann. NY Acad. Sci.*, 1973, 204:416-433.

47. Resnati G, Boldyreva E, Bombicz P, Kawano M. Supramolecular interactions in the solid state. *Int. Un. Crystallogr. Journ.*, 2015, 2(6):675-690.

48. Haketa Y, Takayama M, Maeda H. Solid-state supramolecular assemblies consisting of planar charged species. *Org. Biomol. Chem.*, 2012, 10(13):2603-2606.

49. Borovkov VV, Harada T, Hembury GA, Inoue Y, Kuroda R. Solid-state supramolecular chirogenesis: high optical activity and gradual development of zinc octaethylporphyrin aggregates. *Angew. Chem. Int. Ed.*, 2003, 42(15):1746-1749.

50. Gibb BC. A solid-state supramolecular sweet spot. *Angew. Chem. Int. Ed.*, 2003,42(15):1686-1687.

51. Shiu KB, Lee HC, Lee GH, Ko BT, Wang Y, Lin CC. Solid-state supramolecular organization of supermolecules into a truly molecular zeolite. *Angew. Chem. Int. Ed.*, 2003, 42(26):2999-3001.

52. Cope FW. Solid state theory of competitive diffusion of associated Na+ and K+ in cells by free cation and vacancy (hole) mechanisms, with application to nerve. *Physiol. Chem. Phys.*, 1977, 9(4-5):389-398.

53. Cope FW. The solid-state physics of electron and ion transport in biology. *Adv. Biol. Med. Phys.*, 1970, 13:1-42.

54. Cope FW. Solid state physical mechanisms of biological energy transduction. *Ann. NY Acad. Sci.*, 1974, 227:636-640.

55. Cope FW. Solid state physical replacement of Hodgkin-Huxley theory. Phase transformation kinetics of axonal potassium conductance. *Physiol. Chem. Phys.*, 1977, 9(2):155-160.

56. Nishimura M. [Energy transfer in solid-state and membrane systems in photosynthesis]. Seikagaku. 1968, 40(8):347-356. (Art. in Japan).

57. Brunori M, Santucci R, Campanella L, Tranchida G. Membrane-entrapped microperoxidase as a 'solid-state' promoter in the electrochemistry of soluble metalloproteins. *Biochem. J.*, 1989, 264(1):301-304

58. Bando H, Hisada H, Ishida H, Hata Y, Katakura Y, Kondo A. Isolation of a novel promoter for efficient protein expression by Aspergillus oryzae in solid-state culture. *Appl. Microbiol. Biotechnol.*, 2011, 92(3):561-569.

59. Ishida H, Hata Y, Kawato A, Abe Y. Improvement of the glaB promoter expressed in solid-state fermentation (SSF) of Aspergillus oryzae. *Biosci. Biotech. Biochem.*, 2006, 70(5):1181-1187.

60. Martin CR. Nanomaterials: a membrane-based synthetic approach. *Science*, 1994, 266(5193):1961-1966.

61. Cope FW. Biological sensitivity to weak magnetic fields due to biological superconductive Josephson junctions? *Physiol. Chem. Phys.*, 1973, 5(3):173-176.

62. Cope FW. Superconductivity - a possible mechanism for non-thermal biological effects of microwaves. *J. Microw. Power.*, 1976, 11(3):267-270.

63. Cope FW. On the relativity and uncertainty of distance, time, and energy measurements by man. (1) Derivation of the Weber psychophysical law from the Heisenberg uncertainty principle applied to a superconductive biological detector. (2) The reverse derivation. (3) A human theory of relativity. *Physiol. Chem. Phys.*, 1981, 13(4):305-311.

64. Cope FW. Enhancement by high electric fields of superconduction in organic and biological solids at room temperature and a role in nerve conduction? *Physiol. Chem. Phys.*, 1974, 6(5):405-410.

65. Cope FW. Discontinuous magnetic field effects (Barkhausen noise) in nucleic acids as evidence for room temperature organic superconduction. *Physiol. Chem. Phys.*, 1978, 10(3):233-246.

66. Cope FW. Preliminary studies of magnetic field facilitation of electric conduction in electrically switched "on" dye films that may be room-temperature superconductors. *Physiol. Chem. Phys.,* 1982, 14(5):423-430.

67. Maugh TH New organic superconductor. *Science.* 1984, 226(4670):37.

68. Crabtree GW, Carlson KD, Williams JM. Organic superconductor. *Science*, 1984,

226(4674):494.

69. Wudl F, Nalewajek D, Troup JM, Extine MW. Electron Density Distribution in the Organic Superconductor (TMTSF)2AsF6. *Science*, 1983, 222(4622):415-417.

70. Dunitz JD. Electron Density Distribution in the Organic Superconductor (TMTSF)2AsF6: Fact and Fancy. *Science*, 1985 Apr 19; 228(4697):353-354.

71. Marton JP. Conjectures on superconductivity and cancer. *Physiol. Chem. Phys.*, 1973, 5(3):259-270

72. Alexiou A, Rekkas J. Superconductivity in human body; myth or necessity. *Adv. Exp. Med Biol.*, 2015, 822:53-58.

73. Bystrov VS, Ovtchinnikova GI, Tazieva TR, Soloshenko AN, Pirogov YA, Novik VK. Bioferroelectricity and related problems: Hydrogen-bonded ferroelectric-like systems. *Ferroelectrics*, 2001, 258(1):79-88.

74. Bystrov V, Bystrova N. Bioferroelectricity and optical properties of biological systems. *Adv. Org. Inorg. Opt. Mat.*, 2003, 5122:132-136.

75. Bystrov VS. Models of proton dynamics and superprotonic/ionic conduction in hydrogen-bonded ferroelectrics and related (biological) systems. 1. Soliton dynamics in hydrogen-bonded systems. *Ferroelectrics Lett. Sec.*, 2000, 27(5-6):147-159.

76. Tuszynski JA., Craddock TJA, Carpenter EJ. Bio-ferroelectricity at the nanoscale. *J. Comput. Theor. Nanosci.*, 2008, 5(10):2022-2032.

77. Tokimoto T, Shirane K. Ferroelectric diffused electrical bilayer model for membrane excitation. *Ferroelectrics*, 1993, 141(1):297-305.

78. Leuchtag HR. Fit of the dielectric anomaly of squid axon membrane near heat-block temperature to the ferroelectric Curie-Weiss law. *Biophys. Chem.*, 1995, 53(3):197-205.

79. Tokimoto T Shirane K. Ferroelectric diffused electrical bilayer model for membrane excitation II. Voltage clamped responses. *Ferroelectrics*, 1993, 146(1):73-80.

80. Bdikin I, Bystrov V, Kopyl S, Lopes RPG, Delgadillo I, Gracio J, Mishina E, Sigov A, Kholkin AL. Evidence of ferroelectricity and phase transition in pressed diphenylalanine peptide nanotubes. *Appl. Phys. Lett.*, 2012, 100(4): 043702-1-043702-4.

81. Bystrov VS, Paramonova E, Bdikin I, Kopyl S, Heredia A, Pullar RC, Kholkin AL. Bioferroelectricity: diphenylalanine peptide nanotubes computational modeling and ferroelectric properties at the nanoscale. *Ferroelectrics*, 2012, 440(1):3-24.

82. Bystrov V, Bystrova N, Kisilev D, Paramonova E, Kuhn M, Kleim H, Kholkin A. Molecular model of polarization switching and nanoscale physical properties of thin ferroelectric Langmuir-Blodgett P (VDF-TrFE) films. Integr. *Ferroelectrics*, 2008, 99(1):31-40.

83. Turina AV, Clop PD, Perillo MA. Synaptosomal membrane-based Langmuir-Blodgett films: a platform for studies on γ-aminobutyric acid type A receptor binding properties. *Langmuir*, 2015, 31(5):1792-1801.

84. Pavinatto FJ, Caseli L, Pavinatto A, dos Santos DS Jr, Nobre TM, Zaniquelli ME, Silva HS, Miranda PB, de Oliveira ON Jr. Probing chitosan and phospholipid interactions using Langmuir and Langmuir-Blodgett films as cell membrane models. *Langmuir*, 2007, 23(14):7666-7671.

85. Brosseau CL, Leitch J, Bin X, Chen M, Roscoe SG, Lipkowski J. Electrochemical and PM-IRRAS a glycolipid-containing biomimetic membrane prepared using Langmuir-Blodgett/Langmuir-Schaefer deposition. *Langmuir*, 2008, 24(22):13058-

13067.
86. Hill K, Pénzes CB, Schnöller D, Horváti K, Bosze S, Hudecz F, Keszthelyi T, Kiss E. Characterisation of the membrane affinity of an isoniazide peptide conjugate by tensiometry, atomic force microscopy and sum-frequency vibrational spectroscopy, using a phospholipids Langmuir monolayer model. *Phys. Chem. Chem. Phys.*, 2010, 12(37):11498-11506.
87. Kramarov SO, Dashko YV. Contribution of some relaxation processes to the fracture energy of segnetoelectric materials. *Strength of Materials*, 1987, 19(10):1384-1388.
88. Samoilovich MI, Rinkevich AB, Bovtun V, Belyanin AF, Kempa M, Nuzhnyy D, Tsvetkov MY, Kleshcheva SM. Optical, magnetic, and dielectric properties of opal matrices with intersphere nanocavities filled with crystalline multiferroic, piezoelectric, and segnetoelectric materials. *Russ. J. Gen. Chem.*, 2013, 83(11):2132-2147.
89. Leuchtag HR. Indications of the existence of ferroelectric units in excitable-membrane channels. *J. Theor. Biol.*, 1987. 127(3):321-340.
90. Leuchtag HR. Phase transitions and ion currents in a model ferroelectric channel unit. *J. Theor. Biol.*, 1987, 127(3):341-359.
91. Bystrov VS. Ferroeleciric phason model of sodium channels in biomemberanes. *Ferroelectr. Lett. Sect.*, 1992, 13(6):127-136.
92. Bystrov V, Rolov B, Yurkevich V. Photoferroelectric phenomena in ferroelectrics semiconductors caused by fluctuons and phasons. *Ferroelectrics*, 1984, 55(1):299-302.
93. Bystrov VS, Leuchtag HR. Bioferroelectricity: Modeling the transitions of the sodium channel. *Ferroelectrics,* 1994, 155(1):19-24.
94. Bystrov VS, Lakhno VD, Molchanov M. Ferroelectric active models of ion channels in biomembranes. *J. Theor. Biol.*, 1994, 168(4):383-393.
95. Bystrov VS. Ferroelectric liquid crystal models of ion channels and gating phenomena in biological membranes. *Ferroelectrics Lett. Sect.*, 1997, 23(3-4):87-93.
96. Leuchtag HR. Bioferroelectricity in models of voltage-dependent ion channels. *Ferroelectrics,* 2000, 236(1):23-33.
97. Leuchtag HR, Bystrov VS. Theoretical models of conformational transitions and ion conduction in voltage-dependent ion channels: Bioferroelectricity and superionic conduction. *Ferroelectrics*, 1999, 220(1):157-204.
98. Bystrov VS, Leuchtag H. R. Phase transitions in the ferroelectric-active model of ion channels of biomembranes. *Ferroelectrics,* 1996, 186(1):305-307.
99. Leuchtag HR. Do sodium channels in biological membranes undergo ferroelectric-superionic transitions? *IEEE 7th International Symposium on Applications of Ferroelectrics* (Urbana-Champaign, IL, 1990), p. 279-283.
100. Lomov AA, Shitov NV, Bushuev VA, Baranov AI. Structural phase transition in a surface layer of cesium deuterosulfate single crystals. *JETP Lett.*, 1992, 55(5):296-300.
101. Kirpichnikova LF, Polomska M, Volyak Y, Hilczer B. On the characteristic changes in the domain structure and conductivity of $CsDSO_4$ crystals near the superionic phase transition. *JETP Lett.*, 1996, 63(11):912-916.
102. Wozniak MA, Chen CS. Mechanotransduction in development: a growing role for contractility. *Nat. Rev. Molec. Cell Biol.*, 2009, 10:34-43.
103. Bourgine P, Lesne A. (Eds.) *Morphogenesis: Origins of Patterns and Shapes.* Heidelberg, Springer Science & Business

Media, 2011, 346 p.

104.   Rolov B, Ivin V, Lorencs Y. Physics of ferroelectric liquids. *Ferroelectrics*, 1984, 55(1):159-162.

105.   Klapp S, Forstmann F. Stability of ferroelectric fluid and solid phases in the Stockmayer model. *EPL* (*EuroPhysics Letters*), 1997, 38(9):663-668.

106.   Petschek RG, Wiefling KM. Novel ferroelectric fluids. *Phys. Rev. Lett.*, 1987, 59(3): 343-346.

107.   Clark NA, Lagerwall ST. Physics of ferroelectric fluids: the discovery of a high-speed electro-optic switching process in liquid crystals. *Recent Developments in Condensed Matter Physics*, 1981, 4:309-319.

108.   Cantillon-Murphy P, Wald LL, Adalsteinsson E, Zahn M. Heating in the MRI environment due to superparamagnetic fluid suspensions in a rotating magnetic field. *J. Magn. Magn. Mater.*, 2010, 322(6):727-733.

109.   King LB, Meyer E, Hopkins MA, Hawkett BS, Jain N. Self-assembling array of magnetoelectrostatic jets from the surface of a superparamagnetic ionic liquid. *Langmuir,* 2014, 30(47):14143-14150

110.   Hartmann P, Donkó Z, Rosenberg M, Kalman GJ. Waves in two-dimensional superparamagnetic dusty plasma liquids. *Phys. Rev. E: Stat. Nonlin. Soft Mat. Phys.*, 2014, 89(4):043102-1-043102-9.

111.   Tokimoto T, Shirane K, Kushibe H. Self-organized chemical model and approaches to membrane excitation. *Ferroelectrics*, 1999, 220(1):273-289.

112.   Cope FW. A theory of ion transport across cell surfaces by a process analogous to electron transport across liquid-solid interfaces. *Bull. Math. Biophys.*, 1965, 27:99-109.